\PassOptionsToPackage{unicode}{hyperref}
\PassOptionsToPackage{hyphens}{url}
\PassOptionsToPackage{dvipsnames,svgnames,x11names}{xcolor}
\documentclass[
  10pt,
  a4paper,
]{article}
\usepackage{amsmath,amssymb}
\usepackage{iftex}
\ifPDFTeX
  \usepackage[T1]{fontenc}
  \usepackage[utf8]{inputenc}
  \usepackage{textcomp} 
\else 
  \usepackage{unicode-math} 
  \defaultfontfeatures{Scale=MatchLowercase}
  \defaultfontfeatures[\rmfamily]{Ligatures=TeX,Scale=1}
\fi
\usepackage{lmodern}
\ifPDFTeX\else
\fi
\IfFileExists{upquote.sty}{\usepackage{upquote}}{}
\IfFileExists{microtype.sty}{
  \usepackage[]{microtype}
  \UseMicrotypeSet[protrusion]{basicmath} 
}{}
\makeatletter
\@ifundefined{KOMAClassName}{
  \IfFileExists{parskip.sty}{%
    \usepackage{parskip}
  }{
    \setlength{\parindent}{0pt}
    \setlength{\parskip}{6pt plus 2pt minus 1pt}}
}{
  \KOMAoptions{parskip=half}}
\makeatother
\usepackage{xcolor}
\usepackage[margin=24mm]{geometry}
\usepackage{graphicx}
\makeatletter
\def\maxwidth{\ifdim\Gin@nat@width>\linewidth\linewidth\else\Gin@nat@width\fi}
\def\maxheight{\ifdim\Gin@nat@height>\textheight\textheight\else\Gin@nat@height\fi}
\makeatother
\setkeys{Gin}{width=\maxwidth,height=\maxheight,keepaspectratio}
\makeatletter
\def\fps@figure{htbp}
\makeatother
\usepackage{svg}
\providecommand{\tightlist}{%
  \setlength{\itemsep}{0pt}\setlength{\parskip}{0pt}}
\ifLuaTeX
\usepackage[bidi=basic]{babel}
\else
\usepackage[bidi=default]{babel}
\fi
\babelprovide[main,import]{british}

\def\languageshorthands#1{}
\usepackage[all]{nowidow}
\usepackage{needspace}
\usepackage{tabularx}
\usepackage{array}
\usepackage{booktabs}
\usepackage{seqsplit}
\usepackage{fancyhdr}
\ifLuaTeX
  \usepackage{selnolig}  
\fi
\IfFileExists{bookmark.sty}{\usepackage{bookmark}}{\usepackage{hyperref}}
\IfFileExists{xurl.sty}{\usepackage{xurl}}{} 
\hypersetup{
  pdftitle={Who Assures the Verifier? An Executable Assurance-Locus Audit of the European Digital Identity Wallet},
  pdfauthor={Anton Sokolov --- Tyche Institute --- anton.sokolov@tyche.institute},
  pdflang={en-GB},
  pdfkeywords={European Digital Identity Wallet, relying
party, verifier, conformance testing, assurance
evidence, OpenID4VP, SD-JWT VC},
  colorlinks=true,
  linkcolor={blue},
  filecolor={Maroon},
  citecolor={Blue},
  urlcolor={blue},
  pdfcreator={LaTeX via pandoc}}

\title{Who Assures the Verifier? An Executable Assurance-Locus Audit of
the European Digital Identity Wallet}
\usepackage{etoolbox}
\makeatletter
\providecommand{\subtitle}[1]{
  \apptocmd{\@title}{\par {\large #1 \par}}{}{}
}
\makeatother
\subtitle{Preprint v0.1 --- technical pre-results version --- not peer
reviewed}
\author{Anton Sokolov --- Tyche Institute ---
anton.sokolov@tyche.institute}
\date{2026-08-11}

\begin{document}
\maketitle
\begin{abstract}
The European Digital Identity Wallet (EUDI Wallet) architecture places
material duties on relying parties: they register services and intended
uses, authenticate to Wallet Units, validate presentations and trust
anchors, and make risk-based status decisions. Wallet certification and
the emerging Functional Conformance Assessment Framework provide
increasingly structured wallet-side evidence. A different question
remains: what independently rerunnable evidence shows that the concrete
relying-party verifier version used in a transaction enforced the
applicable request, presentation and reliance-decision controls? We
conduct an assurance-locus audit of current law, Architecture and
Reference Framework (ARF) 3.0.0, ETSI metadata, FCAF scope and three
pinned open-source verifier codebases. We then design a 17-rule research
profile, a machine-readable evidence receipt and 36 frozen synthetic
transactions. Three heterogeneous study-authored implementations
(Python, JavaScript and jq) execute 108 cases with zero oracle
mismatches and zero cross-path disagreements. The experiment establishes
determinism and implementability of the proposed decision model, not
product conformance or certification. Source inspection finds
substantial protocol-verification mechanisms in all three public
codebases but no single audited evidence object joining RP registration
and purpose, exact verifier/policy version, transaction verdict and
downstream attribute use. We therefore propose an
RP-as-system-under-test evidence unit that complements, rather than
displaces, wallet certification, registration and supervision. A
preregistered census of 26 appointed experts is prepared to test content
validity and governance feasibility; recruitment awaits the applicable
ethics/data-protection determination.
\end{abstract}

{
\hypersetup{linkcolor=}
\setcounter{tocdepth}{3}
\tableofcontents
}
\hypertarget{introduction}{%
\section{1. Introduction}\label{introduction}}

European Digital Identity Wallet assurance is often narrated from the
Wallet Unit outward: secure keys, certified components,
privacy-preserving disclosure, interoperable presentation and a
trustworthy wallet lifecycle. That orientation is necessary. It is not
the whole relying transaction. A service ultimately acts because a
relying-party (RP) verifier accepts a presentation and a business
component relies on selected attributes. Assurance about the wallet does
not, by logical implication, establish that the deployed verifier used
the correct trust anchors, bound the presentation to the current
transaction, respected the RP's registered service and intended use, or
limited the later decision to the attributes approved for that
transaction.

This article asks a deliberately bounded question:

\begin{quote}
What independently rerunnable evidence connects a concrete EUDI RP
verifier version to the registration, presentation and reliance-decision
controls applicable to one transaction?
\end{quote}

The question does \textbf{not} assume that relying parties are
unregulated or that existing EUDI documents omit verifier requirements.
Regulation (EU) No 910/2014, as amended by Regulation (EU) 2024/1183,
introduces RP registration duties; Commission Implementing Regulation
(EU) 2025/848 specifies registration mechanics; ETSI TS 119 475
describes RP information supporting the user's authorisation decision;
and ARF 3.0.0 contains explicit RP requirements for signatures, trust
anchors, device binding, status and registration certificates
{[}1--4{]}. The emerging Functional Conformance Assessment Framework
(FCAF) initially treats the Wallet Solution as its System Under Test
(SUT), while explicitly anticipating other future SUTs, including
Relying Parties {[}5{]}.

Our object is consequently not a missing norm but an
\textbf{assurance-locus seam}: normative duties and transaction metadata
exist, yet an evaluator, procurer or supervisor may still lack a small,
portable artefact showing what exact verifier and policy version was
exercised, against which frozen negative case, with which stable reason
code, and how the accepted attributes reached a later service decision.

We make six contributions:

\begin{enumerate}
\def\labelenumi{\arabic{enumi}.}
\tightlist
\item
  a current, actor-specific crosswalk separating legal duties,
  wallet-side checks, RP-side checks, governance evidence and rerunnable
  implementation evidence;
\item
  a 17-rule research profile spanning request, presentation and reliance
  phases;
\item
  a JSON evidence-receipt schema that binds the decision to the RP
  registration identity, verifier and policy versions, input hash and
  stable reason code;
\item
  a public corpus of 36 synthetic transactions and three heterogeneous
  executable paths;
\item
  a pinned, non-vulnerability source audit of three public verifier
  implementations; and
\item
  a preregistered expert-census protocol for content validity and
  governance feasibility.
\end{enumerate}

The result is a proposal for an RP-as-SUT evidence unit. It complements
wallet certification and public registration; it neither grants legal
compliance nor creates a new certification scheme.

\begin{figure}
\centering
\includegraphics[width=1\textwidth,height=\textheight]{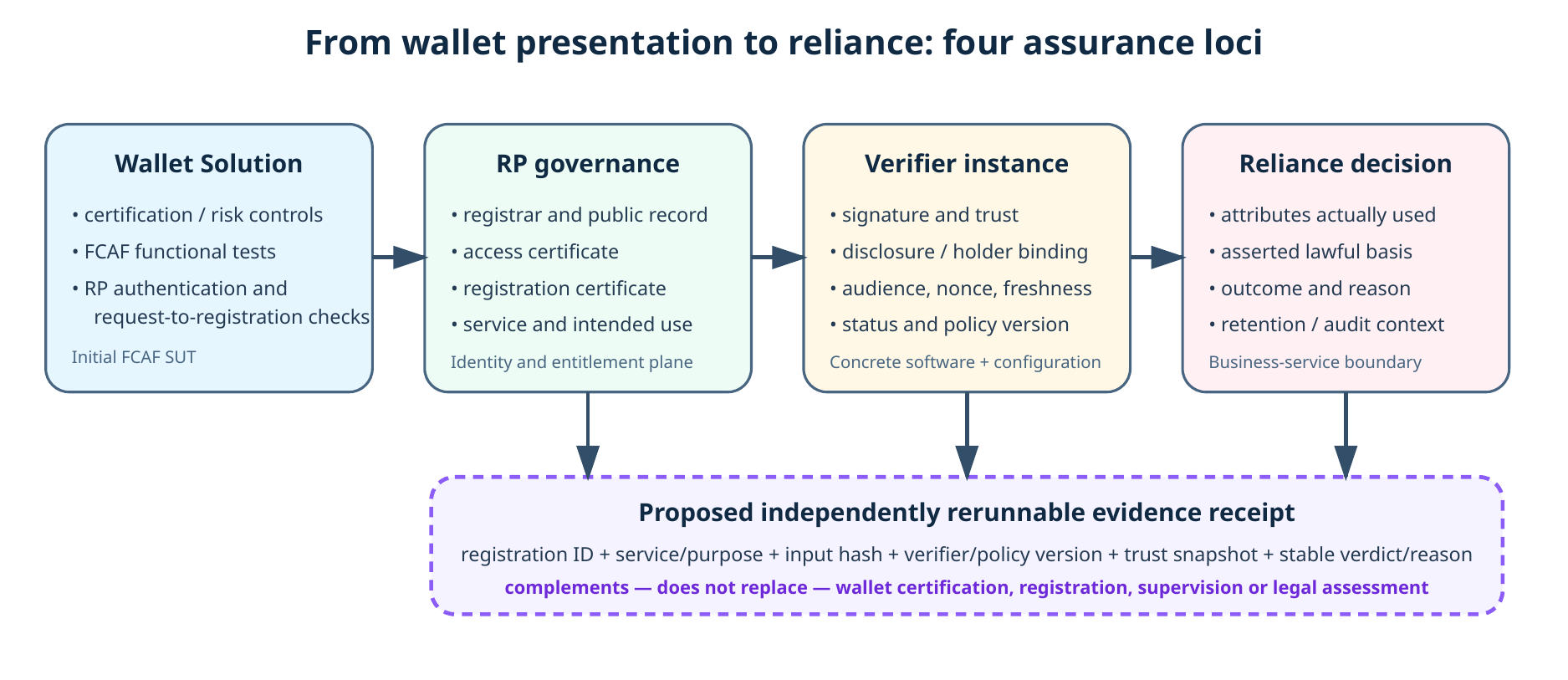}
\caption{Assurance loci around an EUDI Wallet presentation}
\end{figure}

\hypertarget{regulatory-and-technical-setting}{%
\section{2. Regulatory and technical
setting}\label{regulatory-and-technical-setting}}

\hypertarget{registration-and-wallet-side-protection}{%
\subsection{2.1 Registration and wallet-side
protection}\label{registration-and-wallet-side-protection}}

Article 5b of the amended eIDAS Regulation requires a Wallet-relying
party to register in the Member State where it is established and to
state, among other information, the data it intends to request {[}1{]}.
CIR 2025/848 operationalises that registration layer {[}2{]}. ARF 3.0.0
assigns a Registrar to maintain registered RP information, provides for
access and registration certificates, and requires the Wallet Unit to
authenticate an RP Instance and compare requested attributes with the
registration certificate {[}3{]}.

These controls are significant. They create an identity and entitlement
plane before attributes are released. They also primarily protect the
user at the request boundary: the Wallet Unit can identify the RP,
display its registered information, warn about failed checks and limit
or refuse disclosure. An RP Instance must include an applicable
registration certificate in a presentation request. The existence of
these mechanisms is a reason to narrow the research claim, not a reason
to abandon it.

ETSI TS 119 475 specifies RP attributes that support the user's
authorisation decision, including information conveyed through access
and registration certificates {[}4{]}. Such metadata can say who the RP
is and what it registered. It does not by itself demonstrate that the
RP's verifier library, configuration and later business logic correctly
enforced every applicable verification step.

\hypertarget{rp-verification-duties}{%
\subsection{2.2 RP verification duties}\label{rp-verification-duties}}

ARF 3.0.0 contains direct RP obligations. For PID, qualified EAA,
public-body EAA and non-qualified EAA presentations, an RP validates the
applicable signature against the corresponding Trusted List or List of
Trusted Entities. It manages changing trust anchors, should verify
device binding where applicable, and should perform credential-status
checks; deviations from status-checking recommendations require risk
analysis {[}3{]}. OpenID for Verifiable Presentations 1.0 defines
transaction binding through client identifiers, nonce and presentation
processing {[}6{]}. RFC 9901 defines SD-JWT and selective-disclosure
verification mechanics {[}7{]}.

\Needspace{15\baselineskip}

These documents answer ``what should be checked'' at different levels.
They do not necessarily answer the evidence question for one deployment:

\begin{itemize}
\tightlist
\item
  Which exact library, build, dependency set and policy configuration
  executed?
\item
  Which trust-list or LoTE snapshot informed the decision?
\item
  Was the transaction bound to the expected audience and nonce?
\item
  Which registered service and intended use applied?
\item
  Which attributes were approved, and which were actually used
  downstream?
\item
  Can a third party rerun a frozen negative case and obtain the same
  stable refusal reason?
\end{itemize}

Ordinary logs can answer part of this. Conformance suites can answer
another part. Certification can answer a defined target-and-scope
question. The proposed receipt is a join key across these forms of
evidence; it is not a substitute for them.

\hypertarget{certification-and-fcaf-scope}{%
\subsection{2.3 Certification and FCAF
scope}\label{certification-and-fcaf-scope}}

Commission Implementing Regulation (EU) 2024/2981 establishes rules for
EUDI Wallet certification and risk management {[}8{]}. ENISA's draft
candidate EUCC-based scheme, version 0.4.614 on the study date, is
explicitly an early draft and includes wallet-related service-provider
security requirements {[}9{]}. It would therefore be incorrect to say
that the certification work recognises no service-side or RP-originating
risk.

FCAF provides an increasingly structured functional-conformance
framework. Its documentation states that the initial SUT is the Wallet
Solution. The \texttt{WalletSolution\_RelyingParty} test class exercises
the wallet-facing RP interface, but the wallet remains the assessed
system. The methodology lists Relying Parties and corresponding RP test
classes as potential future SUTs {[}5{]}. Our proposal is aligned with
that extensibility: it supplies a small candidate evidence unit for the
RP side without claiming official FCAF status.

\hypertarget{assurance-locus-model}{%
\section{3. Assurance-locus model}\label{assurance-locus-model}}

We distinguish four loci (Figure 1).

\textbf{L1 --- Wallet Solution.} Wallet certification, secure-component
boundaries, risk treatment and FCAF wallet conformance.

\textbf{L2 --- RP governance.} Registration, access certificates,
registration certificates, service and intended-use declarations,
supervision and lifecycle state.

\textbf{L3 --- verifier implementation.} The concrete code and policy
that parses the presentation; verifies signature, disclosure,
holder/device binding, audience, nonce, freshness and status; and emits
a verdict.

\textbf{L4 --- reliance decision.} The application-specific step that
consumes verified attributes, applies asserted lawful basis and policy,
and grants, denies or conditions a service.

The central failure mode is not necessarily a defect within one locus.
It is an \textbf{unjoined evidence chain}. A valid registration at L2
does not prove correct cryptography at L3. A successful cryptographic
result at L3 does not prove that L4 used only the approved attributes or
the intended policy. A certified wallet at L1 does not prove the RP
deployment at L3 or L4.

We define independently rerunnable evidence as an artefact that enables
another authorised evaluator to: (i) identify the profile and
implementation version; (ii) reconstruct or obtain a privacy-safe test
input; (iii) rerun the decision procedure; and (iv) compare a
deterministic verdict and stable reason code. ``Independent'' refers to
the ability to rerun, not necessarily to organisational ownership of the
original test.

\Needspace{16\baselineskip}

\hypertarget{method}{%
\section{4. Method}\label{method}}

\hypertarget{study-design}{%
\subsection{4.1 Study design}\label{study-design}}

The study combines standards engineering and design-and-evaluation
research in five stages:

\begin{enumerate}
\def\labelenumi{\arabic{enumi}.}
\tightlist
\item
  normative crosswalk;
\item
  profile and receipt design;
\item
  frozen negative-test corpus;
\item
  multi-path execution; and
\item
  pinned public-source inspection.
\end{enumerate}

A sixth, human-participant stage is preregistered but not yet opened.
Separating it prevents opinions collected before consent and ethics
determination from being retrofitted into evidence.

\hypertarget{normative-crosswalk}{%
\subsection{4.2 Normative crosswalk}\label{normative-crosswalk}}

We reviewed the consolidated eIDAS text, CIR 2025/848, CIR 2024/2981,
ARF release 3.0.0, ETSI TS 119 475 v1.2.1, FCAF current documentation,
the ENISA candidate scheme page, OpenID4VP 1.0 and RFC 9901 {[}1--9{]}.
Requirements were mapped to responsible actor, normal transaction
evidence, proposed rerunnable evidence and an explicit boundary.

The inclusion rule was functional relevance to one of three phases:
request, presentation or reliance. The crosswalk refuses two inferences:
a requirement is not evidence that a concrete implementation satisfies
it; failure to locate an evidence object in an audited source surface is
not evidence that the ecosystem lacks the control.

\hypertarget{research-profile}{%
\subsection{4.3 Research profile}\label{research-profile}}

The profile contains 17 ordered refusal rules and one acceptance code
(Table 1). Ordering matters when a synthetic transaction violates
multiple conditions: the first applicable rule is the canonical reason.
This makes the oracle deterministic and avoids implementations selecting
different but individually plausible error labels.

Table 1. Research profile.

\begin{center}
\begingroup
\footnotesize
\setlength{\tabcolsep}{4pt}
\renewcommand{\arraystretch}{1.18}
\begin{tabularx}{\textwidth}{@{}>{\raggedright\arraybackslash}p{0.13\textwidth}>{\raggedright\arraybackslash\ttfamily}p{0.31\textwidth}>{\raggedright\arraybackslash}X@{}}
\toprule
\normalfont\textbf{Phase} & \normalfont\textbf{Stable code} & \textbf{Condition} \\
\midrule
Presentation & R\_PARSE\_INVALID & Input is not syntactically processable \\
Request & R\_RP\_UNREGISTERED & RP is not registered \\
Request & R\_RP\_SUSPENDED & Registration is suspended or cancelled \\
Request & R\_RP\_IDENTITY\_MISMATCH & Authenticated and registered RP identities differ \\
Request & R\_PURPOSE\_MISMATCH & Purpose or intended use is not registered \\
Request & R\_ATTRIBUTE\_OVER\_REQUEST & Requested attributes exceed registration \\
Presentation & R\_ISSUER\_SIGNATURE\_INVALID & Issuer authentication fails \\
Presentation & R\_DISCLOSURE\_UNBOUND & Disclosed value is not issuer-bound \\
Presentation & R\_HOLDER\_BINDING\_INVALID & Applicable holder or device binding fails \\
Presentation & R\_AUDIENCE\_MISMATCH & Audience is not the intended verifier \\
Presentation & R\_NONCE\_MISMATCH & Challenge binding fails \\
Presentation & R\_PRESENTATION\_STALE & Presentation exceeds freshness policy \\
Presentation & R\_CREDENTIAL\_INACTIVE & Credential is inactive or revoked \\
Presentation & R\_ISSUER\_UNTRUSTED & Issuer is outside applicable trust policy \\
Reliance & R\_USE\_EXCEEDS\_APPROVAL & Downstream use exceeds approved attributes \\
Reliance & R\_LAWFUL\_BASIS\_UNRECORDED & Asserted lawful basis is not recorded \\
Reliance & R\_AUDIT\_EVIDENCE\_INCOMPLETE & Version, policy, registration or reason evidence is incomplete \\
\bottomrule
\end{tabularx}
\endgroup
\end{center}

The last three rules deliberately extend beyond protocol validity. They
test the join from accepted presentation to reliance. They are research
requirements, not represented as adopted EUDI requirements or
legal-compliance tests.

\Needspace{24\baselineskip}

\hypertarget{evidence-receipt}{%
\subsection{4.4 Evidence receipt}\label{evidence-receipt}}

For every execution, the harness emits a JSON Lines receipt containing:

\begin{itemize}
\tightlist
\item
  schema, profile and profile-version identifiers;
\item
  implementation name;
\item
  vector identifier and canonical SHA-256 input hash;
\item
  verdict and ordered reason codes;
\item
  RP registration identifier;
\item
  verifier and policy versions; and
\item
  frozen evaluation time.
\end{itemize}

A production design would additionally bind trust-list/LoTE snapshots,
software supply-chain attestations and privacy-safe transaction
identifiers. Those fields are omitted from the minimal experiment to
avoid pretending that the synthetic corpus has a live trust
infrastructure.

\hypertarget{corpus}{%
\subsection{4.5 Corpus}\label{corpus}}

The generator produces 36 synthetic, non-personal transactions: six
valid and 30 invalid. Cases cover all 17 rules, boundary combinations
and precedence cases. Six cases vary accepted attribute sets and
purposes; rejection multiplicities range from one to four cases per
reason. Inputs use boolean or finite-state facts (for example
\texttt{issuer\_signature\_valid} and \texttt{credential\_status}) so
the study evaluates the decision profile rather than reimplementing
cryptographic primitives.

This abstraction is intentional and limiting. The corpus can show
whether independently written decision paths implement the same policy.
It cannot establish that a production cryptographic library correctly
validates COSE, JOSE, X.509, SD-JWT or mdoc structures.

\hypertarget{implementations-and-oracle}{%
\subsection{4.6 Implementations and
oracle}\label{implementations-and-oracle}}

We implemented the profile three times:

\begin{itemize}
\tightlist
\item
  imperative Python;
\item
  functional-style JavaScript; and
\item
  declarative jq.
\end{itemize}

Each path receives the identical JSON corpus and emits an ordered result
set. The harness compares every result against the frozen expected
verdict and reason, then checks pairwise agreement across paths. It
emits per-path results, combined receipts, a summary and SHA-256
manifest.

All three paths were authored for this study. Language and
programming-style heterogeneity reduce the chance of one shared
control-flow bug, but do not provide organisational independence.

\hypertarget{pinned-public-source-audit}{%
\subsection{4.7 Pinned public-source
audit}\label{pinned-public-source-audit}}

On 2026-08-11 we pinned three repositories: the EC EUDI verifier
endpoint (\texttt{db544250…}), OpenWallet Foundation Multipaz
(\texttt{570aa247…}) and walt.id identity (\texttt{f773918…})
{[}10--12{]}. Selection was purposive: official reference
implementation, cross-platform open source, and a feature-rich
commercial open-source stack. We inspected public source for mechanisms
corresponding to profile families, not for exploitable defects.

We recorded only source-presence observations. We did not claim the
inspected commit was deployed, did not exercise a hosted service, and
did not map every configuration. Potentially security-relevant leads are
excluded from named findings until exact runtime reproduction and
coordinated disclosure.

\Needspace{15\baselineskip}

\hypertarget{expert-census-stage}{%
\subsection{4.8 Expert-census stage}\label{expert-census-stage}}

The supplied population frame contains 26 appointed experts. The
preregistered design invites all 26 individually and reports both
delivered-invitation and full-frame denominators. The questionnaire asks
respondents to choose an assurance locus, rate the need for rerunnable
evidence, classify the profile controls, select receipt fields and
retest triggers, and identify objections and next steps.

Aggregate use is separately consented; anonymous and named quotation
permissions are optional and default to no. Responses are personal
views, never inferred ENISA or employer positions. Recruitment remains
closed pending an ethics/data-protection determination requested from
the TalTech Academic Ethics Committee on 2026-08-11. No survey results
are reported in this manuscript version.

\hypertarget{results}{%
\section{5. Results}\label{results}}

\hypertarget{normative-locus}{%
\subsection{5.1 Normative locus}\label{normative-locus}}

The crosswalk produced three findings.

First, the RP side is not normatively empty. Registration certificates,
access certificates, intended-use information and Wallet Unit request
checks create a substantial governance-and-user-protection layer. ARF
3.0.0 also directs RP signature, trust-anchor, device-binding and status
behaviour.

Second, wallet conformance and wallet certification have a different
target from the concrete RP verifier deployment. FCAF's initial SUT is
the Wallet Solution, including its interaction with an RP counterparty.
Testing the wallet against an RP interface is not equivalent to treating
the RP implementation as the SUT.

Third, a recurring evidence seam remains between L2, L3 and L4. The
reviewed sources do not define one minimal, portable receipt that binds
registration/service/intended-use context to exact verifier and policy
versions, a replayable input or vector hash, a stable reason-coded
outcome, and the attributes actually consumed by the reliance decision.
This is a finding about the reviewed evidence model, not proof of
universal absence in Member-State or proprietary systems.

\hypertarget{executable-profile}{%
\subsection{5.2 Executable profile}\label{executable-profile}}

Table 2 reports the frozen experiment.

Table 2. Execution summary.

\begin{center}
\begingroup
\small
\setlength{\tabcolsep}{5pt}
\renewcommand{\arraystretch}{1.12}
\begin{tabularx}{\textwidth}{@{}>{\raggedright\arraybackslash}p{0.29\textwidth}>{\raggedright\arraybackslash}X@{}}
\toprule
\textbf{Measure} & \hfill\textbf{Result} \\
\midrule
Synthetic vectors & \hfill 36 \\
Accept / reject vectors & \hfill 6 / 30 \\
Study-authored paths & \hfill 3 \\
Total executions & \hfill 108 \\
Oracle mismatches & \hfill 0 \\
Cross-path disagreements & \hfill 0 \\
Corpus SHA-256 & {\scriptsize\ttfamily\seqsplit{be634b573fc15ca0bbd40b3e68e550e4170be671f8e3dbe3abd8bce69a6553ff}} \\
Profile SHA-256 & {\scriptsize\ttfamily\seqsplit{136bff4e2b46e970e03398e1b02a747f12b5dbabcb58817abaca5cded8c4fbbc}} \\
\bottomrule
\end{tabularx}
\endgroup
\end{center}

All three paths produced the expected ordered reason code for every
vector. The result establishes two modest properties: the decision model
is implementable in substantially different programming styles, and the
frozen profile/corpus pair is deterministic under the supplied oracle.
It does not establish completeness, real-protocol correctness,
resistance to adversarial parsing, production performance or
external-product conformance.

The receipt set makes every result addressable by implementation, input
hash, RP registration identifier, verifier version and policy version.
This allows a reviewer to distinguish ``the library can perform a
check'' from ``this named build and policy produced this verdict for
this frozen input.''

\hypertarget{public-source-audit}{%
\subsection{5.3 Public-source audit}\label{public-source-audit}}

The EC verifier endpoint derives an expected audience from the channel,
supplies expected audience and nonce to validation, constructs a
time-aware challenge predicate, checks advertised algorithms and
validates transaction-data hashes. The repository warns that it is a
development tool rather than a production application {[}10{]}.

Multipaz's current verifier path invokes
\texttt{PresentmentRecord.verify} with an explicit verification time and
consults a trust manager. The repository also contains older/server
verifier code; because module age and deployment reachability were not
resolved, source observations from that surface are not treated as
current-product findings {[}11{]}.

walt.id's OpenID4VP verifier creates a verification context containing
nonce, audience, origins, transaction data and verification time;
verifies advertised algorithms; and applies configurable policies. Its
documented callback configuration for some trust-authority constraints
illustrates both capability and deployment-policy dependence {[}12{]}.

Collectively, the audit falsifies a simplistic version of the thesis:
real open-source stacks already contain substantial verifier logic. It
supports the refined thesis: capability exists at library/policy
surfaces, while the audited artefacts did not provide one common,
independently rerunnable evidence receipt spanning RP registration
context, concrete configuration/version and downstream reliance.

\hypertarget{expert-census}{%
\subsection{5.4 Expert census}\label{expert-census}}

\textbf{Fieldwork not yet open.} This section is locked until the
ethics/data-protection determination, 28-day fieldwork, analysis lock
and consent-compatible coding are complete. The final article will
report invitations, delivery, completions, item denominators,
non-response, role coverage and disconfirming views. It will not use
working-group membership as endorsement.

\hypertarget{discussion}{%
\section{6. Discussion}\label{discussion}}

\hypertarget{what-the-evidence-gap-isand-is-not}{%
\subsection{6.1 What the evidence gap is---and is
not}\label{what-the-evidence-gap-isand-is-not}}

The evidence gap is not ``nobody assures relying parties.''
Registration, supervision, ordinary application security, procurement,
managed services, protocol conformance and organisation-specific audits
may all contribute. The gap is the absence, in the reviewed common
evidence surfaces, of a minimal join across the version that requested
attributes, the version that verified them, and the policy that relied
on them.

That distinction matters for governance. A new heavyweight certification
programme may be unnecessary or disproportionate. A small RP-as-SUT
conformance unit could instead be attached to existing loci: FCAF,
procurement, national registration, managed verifier services, sector
rules or supervised high-impact deployments. The receipt is deliberately
neutral about which institution owns the assessment.

\hypertarget{why-stable-refusal-reasons-matter}{%
\subsection{6.2 Why stable refusal reasons
matter}\label{why-stable-refusal-reasons-matter}}

A binary ``valid'' result is insufficient for independent review. Stable
reason codes make negative tests comparable across versions,
implementations and time. They also expose policy precedence. For
example, a suspended RP should be rejected before a verifier spends
effort on presentation cryptography; an over-request should be
distinguishable from an invalid credential; a valid presentation should
still fail if downstream use exceeds the approved attribute set.

Stable reasons create risks. They can leak implementation detail, become
brittle API commitments, or encourage checkbox compliance. A deployable
profile should therefore separate public coarse codes from restricted
diagnostic detail and must define privacy-preserving receipt access.

\hypertarget{joining-protocol-and-purpose}{%
\subsection{6.3 Joining protocol and
purpose}\label{joining-protocol-and-purpose}}

Protocol verification and purpose limitation are often owned by
different teams. The proposed profile makes the seam explicit.
Request-phase controls bind registered identity, service, intended use
and attribute set. Presentation-phase controls validate cryptographic
and transaction semantics. Reliance-phase controls compare approved and
used attributes and record the asserted lawful-basis field separately
from wallet-user approval.

Recording a lawful-basis assertion does not make processing lawful. It
prevents one narrower error: treating user approval in a wallet UI as
automatic proof of a controller's legal basis. Legal assessment remains
outside the profile.

\Needspace{16\baselineskip}

\hypertarget{integration-with-fcaf}{%
\subsection{6.4 Integration with FCAF}\label{integration-with-fcaf}}

FCAF already uses a hierarchical SUT/class/layer/area/group/unit
structure and anticipates RP SUTs.

\Needspace{14\baselineskip}

The smallest practical integration would be an
\texttt{RelyingParty\_WalletSolution} security-mechanism unit with:

\begin{enumerate}
\def\labelenumi{\arabic{enumi}.}
\tightlist
\item
  a normative traceability record;
\item
  frozen positive and negative inputs;
\item
  a required stable verdict taxonomy;
\item
  a signed implementation/policy identity;
\item
  a privacy-safe receipt; and
\item
  declared retest triggers.
\end{enumerate}

The proposed corpus is a seed, not a ready official test book.
Production cases must use actual OpenID4VP, SD-JWT VC and mdoc objects;
include malformed and adversarial encodings; and define trust/status
fixtures.

\hypertarget{deployment-and-privacy}{%
\subsection{6.5 Deployment and privacy}\label{deployment-and-privacy}}

Rerunnable evidence should not become a second personal-data ledger. A
production receipt should prefer hashes, short-lived pseudonymous
transaction references and versioned public policy identifiers. Raw
presentations should remain under the RP's existing legal and security
controls. Access to a replay bundle may need to be restricted to
accredited laboratories, supervisors or incident investigators. Public
reproducibility can use synthetic vectors, while transaction-specific
auditability uses privacy-preserving commitments.

\hypertarget{validity-and-limitations}{%
\section{7. Validity and limitations}\label{validity-and-limitations}}

\textbf{Construct validity.} The 17 rules operationalise a research
concept of verifier assurance. Three reliance-phase rules are not
claimed as adopted EUDI conformance requirements. The expert census is
designed to test content validity rather than to manufacture consensus.

\textbf{Internal validity.} A generator, oracle and three
implementations could share the same conceptual mistake. Ordered rules
and cross-language implementations reduce coding inconsistency but
cannot prove the oracle is normatively correct. Future work should add
externally authored vectors and mutation testing.

\textbf{External validity.} Synthetic booleans do not model production
parsers, cryptography, trust discovery, status privacy, network failures
or business workflows. The source-audit sample is small and purposive.
No inspected source is asserted to equal a particular deployed service.

\textbf{Temporal validity.} EUDI documents are evolving. Findings are
pinned to ARF 3.0.0, the FCAF and ENISA scheme pages as observed on
2026-08-11, and named repository commits. Later baselines may directly
fill the proposed seam.

\textbf{Human evidence.} The 26-person frame is institutionally salient
but not representative of all Member States, RPs, implementers, civil
society or users. Non-response and current-role changes may be
substantial. The study will report denominators and avoid population
inference. A later field study should examine whether real service
decisions use only attributes approved in the wallet transaction; that
work will require substantially stronger data-protection design and
cannot be inferred from protocol traces alone.

\textbf{Security disclosure.} Source inspection can reveal
implementation weaknesses. No named vulnerability claim is made.
Runtime-reproducible issues require private maintainer contact and a
disclosure window before publication.

\hypertarget{future-evaluation}{%
\section{8. Future evaluation}\label{future-evaluation}}

The next technical stage is an externally owned verifier plugfest. At
least three independent products should run the same signed corpus
through documented adapters. Results should distinguish protocol
capability, configured policy and deployed reachability. Mutation
testing should measure whether the suite detects deliberately introduced
faults.

The next governance stage is the preregistered expert census. Its most
useful outcome may be disagreement: laboratories may prefer evaluable
control families, deployers may reject expensive receipts, privacy
experts may constrain transaction evidence, and registrars may resist
owning software assurance. Those tensions should determine placement and
proportionality of any RP-as-SUT unit.

\Needspace{14\baselineskip}

\hypertarget{conclusion}{%
\section{9. Conclusion}\label{conclusion}}

EUDI Wallet relying parties already face meaningful registration and
verification requirements, and public verifier implementations already
contain substantial protocol logic. The assurance question is therefore
not whether the verifier exists in the architecture. It is whether the
behaviour of the exact verifier and policy version can be independently
rerun and connected to the registered purpose and later reliance
decision.

The proposed 17-rule profile, 36-vector corpus and reason-coded receipt
make that question executable. Three study-authored paths agree across
108 runs, showing deterministic implementability while leaving product
conformance unproven. A pinned source audit further narrows the claim:
existing capabilities are real, but the reviewed common artefacts do not
yet join governance, implementation and reliance evidence into one
portable object.

An RP-as-SUT evidence unit is consequently a plausible complement to
wallet certification, RP registration and FCAF---not a replacement for
any of them. Its institutional owner, proportionality, privacy model and
minimum control set remain empirical questions. The preregistered expert
census is designed to answer those questions without converting
appointments into endorsements.

\hypertarget{declarations}{%
\section{Declarations}\label{declarations}}

\hypertarget{ethics}{%
\subsection{Ethics}\label{ethics}}

No human-participant response is reported. A determination request was
sent to the TalTech Academic Ethics Committee on 2026-08-11. Recruitment
will begin only after the applicable ethics and data-protection route is
confirmed.

\hypertarget{data-and-code-availability}{%
\subsection{Data and code
availability}\label{data-and-code-availability}}

The research profile, receipt schema, corpus generator, frozen corpus,
three implementations, receipts, results and SHA-256 manifest are
included in the accompanying repository under
\path{papers/eudi-rp-assurance-locus/artifact/}. The expert protocol,
consent text, instrument and codebook are included under \path{survey/}.
Human response data, if collected, will be shared only to the extent
permitted by consent, privacy obligations and the final determination.

\hypertarget{funding}{%
\subsection{Funding}\label{funding}}

No external funding was received for this study.

\hypertarget{competing-interests}{%
\subsection{Competing interests}\label{competing-interests}}

The author declares no competing interests. Tyche Institute is the
author's research affiliation and is not a certification body for the
EUDI Wallet ecosystem.

\hypertarget{author-contributions}{%
\subsection{Author contributions}\label{author-contributions}}

Anton Sokolov: conceptualisation, methodology, software, investigation,
data curation, writing and project administration. AI-assisted drafting
and coding were used under author supervision; the author remains
responsible for factual verification, analysis and the submitted text.
No contacted expert is an author or endorser by virtue of participation.

\clearpage

\hypertarget{references}{%
\section{References}\label{references}}

\begin{enumerate}
\def\labelenumi{\arabic{enumi}.}
\tightlist
\item
  European Parliament and Council. Regulation (EU) No 910/2014 on
  electronic identification and trust services, consolidated text
  including Regulation (EU) 2024/1183.
  \href{https://eur-lex.europa.eu/legal-content/EN/TXT/?uri=CELEX:02014R0910-20241018}{EUR-Lex
  consolidated text}.
\item
  European Commission. Commission Implementing Regulation (EU) 2025/848
  of 6 May 2025 as regards the registration of wallet-relying parties.
  \href{https://eur-lex.europa.eu/legal-content/EN/TXT/?uri=CELEX:32025R0848}{EUR-Lex}.
\item
  European Commission. European Digital Identity Wallet Architecture and
  Reference Framework, release 3.0.0. 2026.
  \href{https://github.com/eu-digital-identity-wallet/eudi-doc-architecture-and-reference-framework/releases/tag/v3.0.0}{Release
  v3.0.0}.
\item
  ETSI. TS 119 475 V1.2.1: Relying party attributes supporting EUDI
  Wallet user's authorisation decisions. 2026.
  \href{https://www.etsi.org/deliver/etsi_ts/119400_119499/119475/01.02.01_60/ts_119475v010201p.pdf}{ETSI
  PDF}.
\item
  European Commission. Functional Conformance Assessment Framework for
  the European Digital Identity Wallet ecosystem. Accessed 2026-08-11.
  \href{https://conformance.eudi.dev/}{FCAF site}.
\item
  OpenID Foundation. OpenID for Verifiable Presentations 1.0. 2025.
  \href{https://openid.net/specs/openid-4-verifiable-presentations-1_0.html}{OpenID
  specification}.
\item
  Fett D, Yasuda K, Campbell B. RFC 9901: Selective Disclosure for JSON
  Web Tokens. IETF; 2025.
  \href{https://www.rfc-editor.org/rfc/rfc9901.html}{RFC 9901}.
\item
  European Commission. Commission Implementing Regulation (EU) 2024/2981
  laying down rules for the application of Regulation (EU) No 910/2014
  as regards the certification of European Digital Identity Wallets.
  \href{https://eur-lex.europa.eu/legal-content/EN/TXT/?uri=CELEX:32024R2981}{EUR-Lex}.
\item
  ENISA. Draft candidate EU Digital Identity Wallet certification scheme
  v0.4.614 for public review. 2026.
  \href{https://certification.enisa.europa.eu/publications/draft-candidate-eudiw-scheme-v04614-public-review_en}{ENISA
  publication page}.
\item
  European Commission. EUDI verifier endpoint, commit
  \href{https://github.com/eu-digital-identity-wallet/eudi-srv-verifier-endpoint/tree/db5442501ea06907e614377a20d802748e8bfddb}{\texttt{db5442501ea0}}.
  2026.
\item
  OpenWallet Foundation. Multipaz, commit
  \href{https://github.com/openwallet-foundation/multipaz/tree/570aa2475bd5b7e437d9041bf8ff1127bcf86cfb}{\texttt{570aa2475bd5}}.
  2026.
\item
  walt.id. waltid-identity, commit
  \href{https://github.com/walt-id/waltid-identity/tree/f773918a3ad226ba7c0908d58941f18a3b89591d}{\texttt{f773918a3ad2}}.
  2026.
\end{enumerate}

\end{document}